\documentclass[iop]{emulateapj}

\usepackage[colorlinks=true,linkcolor=blue,
  anchorcolor=blue,citecolor=blue,urlcolor=blue]{hyperref}  
\hypersetup{pdfauthor={Eunbin Kim}, 
  pdftitle={Star Formation Activity of Barred Spiral Galaxies}}

\slugcomment{Last updated: \today}
\shorttitle{Star Formation Activity of Barred Spiral Galaxies}
\shortauthors{Kim et al.}

\usepackage{natbib}
\usepackage{subfigure}
\usepackage{url}

\usepackage{verbatim}
\usepackage{color}
\usepackage[usenames,dvipsnames]{xcolor}
\definecolor{green}{rgb}{0.0, 0.4, 0.0}
\definecolor{forestgreen(web)}{rgb}{0.13, 0.55, 0.13}
\definecolor{green(web)}{rgb}{0.13, 0.55, 0.13}
\definecolor{green}{rgb}{0.0, 0.4, 0.0}

\makeatletter

\newcommand{\Rmnum}[1]{\expandafter\@slowromancap\romannumeral #1@}
\makeatother

\begin{document}

\title{STAR FORMATION ACTIVITY OF BARRED SPIRAL GALAXIES}

\author{Eunbin Kim\altaffilmark{1}, Ho Seong Hwang\altaffilmark{2}, Haeun Chung\altaffilmark{3,4}, Gwang-Ho Lee\altaffilmark{3}, Changbom Park\altaffilmark{4}, 
Bernardo Cervantes Sodi\altaffilmark{5}, Sungsoo S. Kim\altaffilmark{1,6}}
\altaffiltext{1}{School of Space Research, Kyung Hee University, Yongin, Gyeonggi 17104, Korea; ebkim@khu.ac.kr}
\altaffiltext{2}{Quantum Universe Center, Korea Institute for Advanced Study, 
  85 Hoegiro, Dongdaemun-gu, Seoul 02455, Korea}  
\altaffiltext{3}{Astronomy Program, Department of Physics and Astronomy, Seoul National University,
1 Gwanak-ro, Gwanak-gu, Seoul 08826, Korea}
\altaffiltext{4}{School of Physics, Korea Institute for Advanced Study, 
  85 Hoegiro, Dongdaemun-gu, Seoul 02455, Korea}  
\altaffiltext{5}{Instituto de Radioastronom\'ia y Astrof\'isica, Universidad Nacional Aut\'onoma de M\'exico, Campus Morelia, A.P. 3-72, C.P. 58089 Michoac\'an, M\'exico}
\altaffiltext{6}{Department of Astronomy \& Space Science, Kyung Hee University, Yongin, Gyeonggi 17104, Korea}

\begin{abstract}
We study the star formation activity of nearby galaxies with bars 
  using a sample of late-type galaxies 
  at 0.02$\leq z \leq$ 0.05489 and $M_r <-19.5$
  from the Sloan Digital Sky Survey. 
We compare the physical properties of strongly and weakly barred galaxies 
  with those of non-barred galaxies that
  have stellar mass and redshift distributions similar to barred galaxies.
We find that the star formation activity of strongly barred galaxies
  probed by starburstiness, $\it{g-r}$, NUV$-r$, and mid-infrared [3.4]$-$[12] colors is,
  on average, lower than that of non-barred galaxies.
However, weakly barred galaxies do not show such a difference 
  between barred and non-barred galaxies.
The amounts of atomic and molecular gas in strongly barred galaxies
  are smaller than those of non-barred galaxies, and
  the gas metallicity is higher in strongly barred galaxies
  than in non-barred galaxies.
The gas properties of weakly barred galaxies again show no difference 
  from those of non-barred galaxies.
We stack the optical spectra of barred and non-barred galaxies
  in several mass bins and fit to the stacked spectra
  with a spectral fitting code, STARLIGHT.  
We find no significant difference in stellar populations 
  between barred and non-barred galaxies
  for both strongly and weakly barred galaxies.
Our results are consistent with the idea
  that the star formation activity of barred galaxies
  is enhanced in the past along with significant gas consumption, 
  and is currently lower than or similar to
  that of non-barred galaxies.
The past star formation enhancement depends on the strength of bars.
\end{abstract}


\section{INTRODUCTION}
Star formation activity of galaxies is strongly affected by 
  both internal and external physical processes.
In the early universe, galaxy evolution is primarily determined by external effects
  that include hierarchical clustering and merging.
Then, galaxy evolution becomes mainly secular, 
  which is a rearrangement of energy and mass 
  by non-axisymmetric galactic inner structures \citep{kormendy04}. 
The non-axisymmetric potential in galaxies 
  can be driven by bars or ovals,
  which can cause the gas to lose angular momentum and 
  to infall into a central region of galaxies from galactic disks.
The gas accumulated in the galactic center
  becomes fuels for the central star formation \citep{athan92}.

Early studies showed that bars can play such a role
  in transporting gas into the galactic center.
When a galaxy contains a bar in the central region, 
  gas far from resonances 
  tends to settle on periodic orbits including $x_1$ orbits
  that are elongated along the major axis of bars 
  \citep{contopoulos80,binney91,morris96}. 
The gas then goes through shocks, 
  flows inwards by losing angular momentum, and 
  moves to the $x_2$ orbits elongated along the minor axis. 
The gas finally settles into stable orbits,
  which are close to the position of the inner Lindblad resonance (ILR) 
  of the bar \citep{simkin80,combes85}.
Because the gas accumulated in the central region
  is used for fuel of central star formation,
  the enhanced star formation activity in the central region 
  could be a good indicator of recent gas inflow into the center \citep{knapen95}.
Many numerical simulations indeed demonstrated that the star formation activity
  can be enhanced in central regions of galaxies when there are galactic bars
  \citep{shlosman90, athan94, combes01,kimss11,kimwt12,seo13,shin17}.

Observations also showed enhanced star formation activity 
  in central regions of barred galaxies.  
For example, \citet{heckman80} found that recent star formation activity
  is more frequently observed in barred galaxies than in non-barred galaxies.  
The different star formation activity between barred and non-barred galaxies
  is also supported by multiwavelength observations
  \citep{hawarden86,devereux87,himmel90,regan06,wang12,lin17}.
The effects of bars tend to be stronger for the galaxies with 
  earlier morphological types \citep{ho97,oh12}.
  
On the other hand, some observations found
  no increase in star formation activity in barred galaxies
  compared to non-barred galaxies
  \citep{pompea90, martinet97, chapelon99, cheung13,willett15}.
Similarly, the amount of gas that is fuel for star formation
  is not larger in barred galaxies than in non-barred galaxies;
  the bar fraction decreases with increasing HI gas fraction of galaxies
  \citep{masters12,sodi17}.
\citet{saintonge12} also suggested that 
  bar instabilities do not significantly affect the star formation budget
  for local galaxies.

The complicated situation is not much different 
  for the role of bars in triggering nuclear activity of galaxies.
\cite{oh12} found that the effects of bars on the activity in galactic nuclei
  are stronger in bluer galaxies than in redder galaxies.
However, \cite{lee12b} showed that
  the nuclear activity of barred galaxies does not differ
  from that of non-barred galaxies
  once the physical properties of host galaxies are well matched
  (see also \citealt{cisternas13,cheung15}).

To better understand the role of galactic bars in triggering star formation activity,
  we compare various physical properties of barred galaxies
  with those of carefully selected control sample of non-barred galaxies.
The various physical parameters 
  representing the star formation activity include starburstiness 
  (a measure of the excess in specific star formation rate of a galaxy
  compared to the specific star formation rate of a main sequence star-forming galaxy 
  with the same mass, \citealt{elbaz11}),
  multiwavelength photometric data including ultraviolet (UV) and mid-infrared,
  the amounts of atomic and molecular gas, and the gas metallicity. 
We compare the star formation activity
  not only between barred and non-barred galaxies, but also
  between strongly and weakly barred galaxies.
We also fit to the optical spectra with STARLIGHT
  to make a detailed comparison of the stellar populations 
  between barred and non-barred galaxies.

Section \ref{data} describes the sample of barred galaxies and its control sample,
  and explains the observational data we use.
We compare the star formation activity
  between barred and non-barred galaxies in Section \ref{results}. 
We discuss the results and conclude in Sections \ref{discuss} and \ref{sum}, respectively.
Throughout, we adopt flat $\Lambda$ cold dark matter cosmological parameters:
  $H_0 = 70$ km s$^{-1}$ Mpc$^{-1}$, 
  $\Omega_{\Lambda}=0.7$, and $\Omega_{m}=0.3$.

\section{DATA}\label{data}
\subsection{Samples of Galaxies with and without Bars}

We use the samples of barred and non-barred galaxies 
  described in \cite{lee12a}, which were constructed
  from a volume-limited sample of 33,391 galaxies
  at 0.02 $\leq z \leq$ 0.05489 and $M_r <  -19.5$\footnote{The $r$-band absolute magnitude, 
  $M_r$, is based on $H_0 = 100$ km s$^{-1}$ Mpc$^{-1}$ in \citet{lee12a}.}
  from the Sloan Digital Sky Survey data release 7
  (SDSS DR7, \citealt{abazajian09}).
\cite{lee12a} identified bars in galaxies 
  through visual inspection of SDSS color images.
They classified barred galaxies into three types based on the relative size of bars
 (i.e. strong, weak and ambiguous), 
 which agrees well with the classification result of \cite{nair10}.
\cite{lee12a} also provide a sample of galaxies without bars.

The catalog contains both early- and late-type galaxies 
  with an axis ratio $b/a >$ 0.6 where bar classification is reliable. 
Galaxy morphology is adopted from 
  the Korea Institute for Advanced Study 
  Value-Added Galaxy Catalog (KIAS VAGC; \citealt{choi10}).
We use only the sample of 10,674 late-type galaxies 
  with strong, weak and no bars in this study:
  2542 strongly-barred (23.8\%), 698 weakly-barred (6.5\%) and
  7434 non-barred (69.7\%) galaxies.
  
Because we are interested in star formation activity of barred galaxies,
  we removed galaxies with active galactic nuclei (AGNs).
These include optical AGNs
  identified with the criteria of \citet{kewley06} 
  based on Baldwin-Phillips-Terlevich (BPT) emission-line ratio diagrams.
We also identified AGNs using the 
  {\it Wide-field Infrared Survey Explorer} ($\it{WISE}$, \citealt{wright10})
  mid-infrared color-color selection 
  criteria of \cite{jarrett11} plus \citet{mateos12},
  and removed them from the sample.

\begin{figure*}
\center
\includegraphics[width=0.95\textwidth]{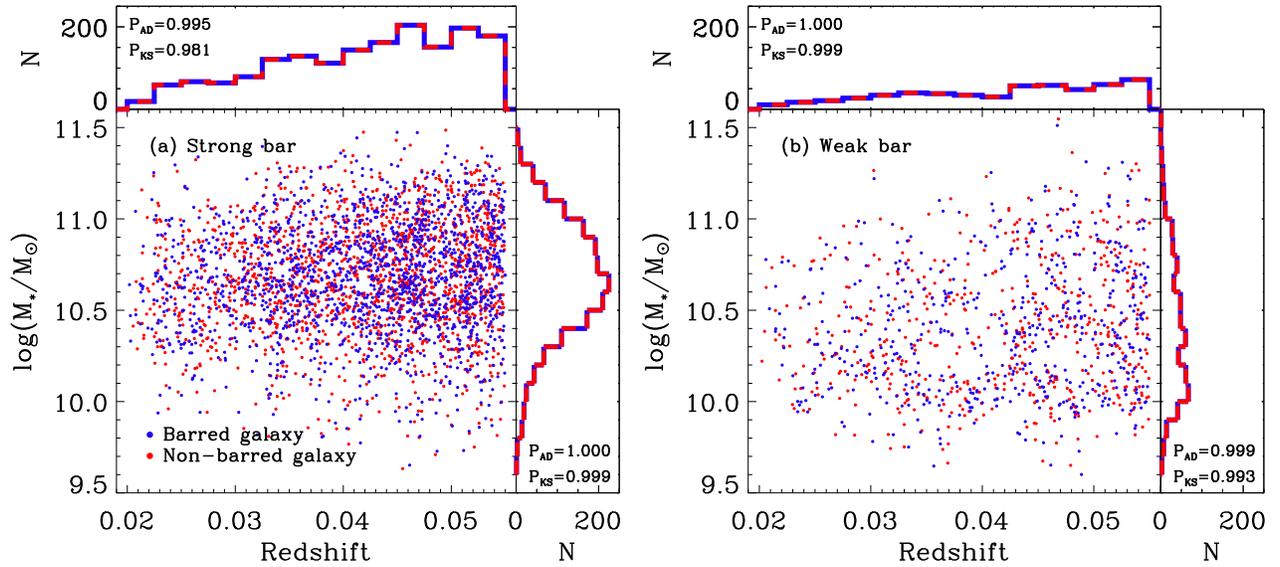}
\caption{Stellar mass as a function of redshift
  for the galaxies with strong (left) and weak (right) bars. 
Blue and red dots represent barred and non-barred galaxies, respectively.
The red histograms are the redshift and stellar mass distributions for barred galaxies,
  which overlap with those distributions of non-barred galaxies (blue histogram). 
Two numbers in each panel indicate $p$-values from the K-S and A-D k-sample tests 
  on the distributions of barred and non-barred galaxies.
}
\label{fig-sam}
\end{figure*}

We finally have a sample of 1686 strongly and 547 weakly barred galaxies 
  at 0.02 $\leq z \leq$ 0.05489 and $M_r < -19.5$.
To compare the physical properties of these barred galaxies
  with those of non-barred galaxies,
  we construct the control sample of barred galaxies 
  using the non-barred galaxies.
To have an unbiased control sample,
  we match the stellar mass and redshift distributions
  of barred and non-barred galaxies; we randomly select
  galaxies from the sample of non-barred galaxies to have
  the same distributions of stellar mass and redshift as
  for the sample of barred galaxies.
We construct the control samples for strongly and weakly barred galaxies separately.  
Figure \ref{fig-sam} shows stellar masses of
  strongly (left panels) and weakly (right panels) barred galaxies 
  as a function of redshift with their control samples.
We examine the redshift and stellar mass distributions 
  of barred and non-barred galaxies using 
  the Kolmogorov-Smirnov (K-S) test and the Anderson-Darling (A-D) k-sample test.
Both tests cannot reject the null hypothesis that 
  the distributions of barred and non-barred galaxies 
  are extracted from the same parent population.
We also examine the distributions of axial ratio and apparent isophotal size 
  of barred and non-barred galaxies that can affect the measurement of star formation activity, 
  and again find no systematic difference between the two samples.

\subsection{Physical Parameters of Galaxies}

The physical parameters of galaxies that we consider in this study
  are star formation rate (SFR), stellar mass, 
  UV/mid-infrared photometric data, atomic and molecular gas masses,
  and gas metallicity.
Here we briefly describe these parameters.

The SFRs of galaxies are adopted from
  the MPA/JHU DR7 VAGC \citep{brinchmann04},
  which provides extinction and aperture corrected
  SFR estimates for SDSS galaxies.
When the SFRs of galaxies cannot be directly 
  measured from the emission lines
  (e.g., AGN and composite galaxies),
  they use the 4000~\AA~break (D4000) to measure SFRs
  (see \citealt{brinchmann04} and the web site$\footnote{
  http://www.mpa-garching.mpg.de/SDSS/DR7/sfrs.html}$ for details).
The stellar mass estimates are also from the MPA/JHU DR7 VAGC,
  which are based on the fit of SDSS five-band photometry 
  with the model of \citet{bc03} (see also \citealt{kauffmann03}).
We convert SFR and stellar mass estimates in the MPA/JHU DR7 VAGC
  that are based on Kroupa initial mass function (IMF, \citealt{kroupa01})
  to those with Salpeter IMF \citep{salpeter55} 
  by dividing them by a factor of 0.7 \citep{elbaz07}.
The gas metallicity (i.e. gas-phase oxygen abundance) is also adopted
  from the MPA/JHU DR7 VAGC \citep{tremonti04}.

We used the multiwavelength photometric data of SDSS galaxies
  compiled in \citet{hwang13}.
The near-UV data are taken from 
  the {\it Galaxy Evolution Explorer} 
  ({\it GALEX}, \citealt{martin05}) general release 6 (GR6),
  which provides a cross-matched table
  (\textbf{xSDSSDR7}) against the SDSS DR7.
We also include the mid-infrared data from the {\it WISE} all-sky survey catalog
  \citep{wright10}.
The catalog provides uniform photometric data for over 747 million objects
  at four mid-infrared bands (3.4, 4.6, 12, and 22 $\micron$m).
The matching tolerance between the SDSS and {\it WISE} objects is 3\arcsec.
We adopt the point source profile-fitting magnitudes, and 
  use only the flux density with the signal-to-noise ratio ${\rm S/N\geq3}$ at each band.

The atomic gas mass, $M_{\rm HI}$, is adopted from 
  the 70 per cent data ($\alpha$.70) of
  the Arecibo Legacy Fast ALFA Survey (ALFALFA, \citealt{giovanelli05,haynes11}),
  which provides HI data for 25,535 galaxies at $z<0.06$.
The molecular gas mass, $M_{\rm H_2}$, is collected 
  from the CO Legacy Database for GASS (COLDGASS, \citealt{saintonge11}) and
  the Atacama Pathfinder Experiment (APEX) 
  Low-redshift Legacy Survey for MOlecular Gas 
 (ALLSMOG, \citealt{bothwell14}).
Among the 1686 strongly barred galaxies 
  (their control sample has the same number of galaxies),
  there are 318 barred and 397 non-barred galaxies with HI detections,
  and are 11 barred and 23 non-barred galaxies with H$_2$ detections.
Among the 547 weakly barred galaxies,
  there are 149 barred and 135 non-barred galaxies with HI detections,
  and are 3 barred and 9 non-barred galaxies with H$_2$ detections.
\begin{table*}[t]
\centering
\caption{Number and Fraction of Galaxies in Each Group}
\begin{tabular}{cccccccc}
\hline
Type &  Strongly Barred Galaxies & & & Weakly Barred Galaxies  & \\
\cline{2-3} \cline{5-6}
       &  Bars & No Bars & & Bars & No Bars  \\
\hline
Starburst &   $~~~3$ ($~0.2\pm$0.1\%)  &   $~~11$ ($~0.7\pm$0.1\%) &&   $~~4$ ($~0.7\pm$0.1\%) & $~10$ ($~1.8\pm$0.1\%) \\
Main Sequence & 1065 (63.1$\pm$0.9\%) & 1139 (67.5$\pm$0.9\%) && 441 (80.6$\pm$1.5\%) & 430 (78.6$\pm$1.5\%) \\
Quiescent & $~618$ (36.7$\pm$0.7\%) & $~536$ (31.8$\pm$0.7\%) && 102 (14.7$\pm$0.7\%) & 107 (19.6$\pm$0.8\%) \\
\hline
Total & 1686 & 1686 && 547 & 547 \\
\hline
\end{tabular}
\label{tab:frac}
\end{table*}
\begin{figure*}
\center
\includegraphics[width=0.7\textwidth]{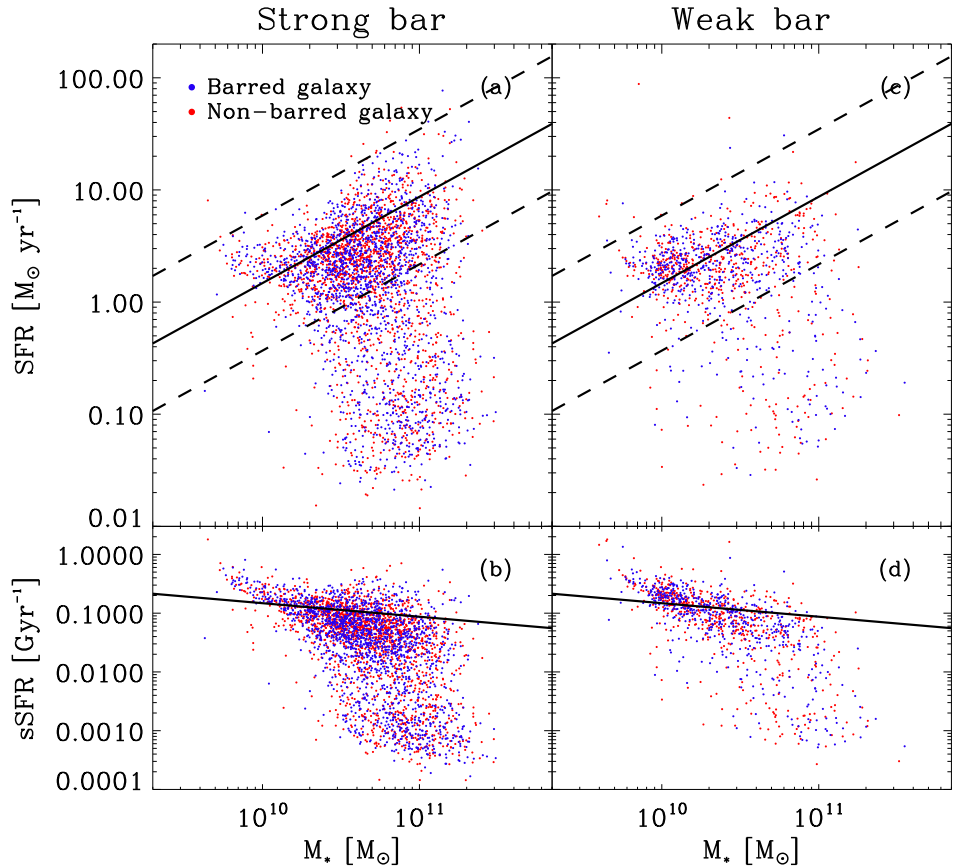}
\caption{SFRs (left top) and sSFRs (left bottom) of galaxies as a function of 
  stellar mass for strongly barred galaxies. 
Blue and red dots represent barred and non-barred galaxies, respectively. 
Solid lines in top and bottom panels are 
  the best fits of the main sequence star-forming galaxies in SDSS \citep{elbaz07},
  and the upper and lower dashed lines are a factor 4 above and below this fit.
(Right) Same as left panels, but for weakly barred galaxies and their control sample.
}\label{fig-sfrm}
\end{figure*}

\begin{figure*}[t]
\center
\includegraphics[width=0.95\textwidth]{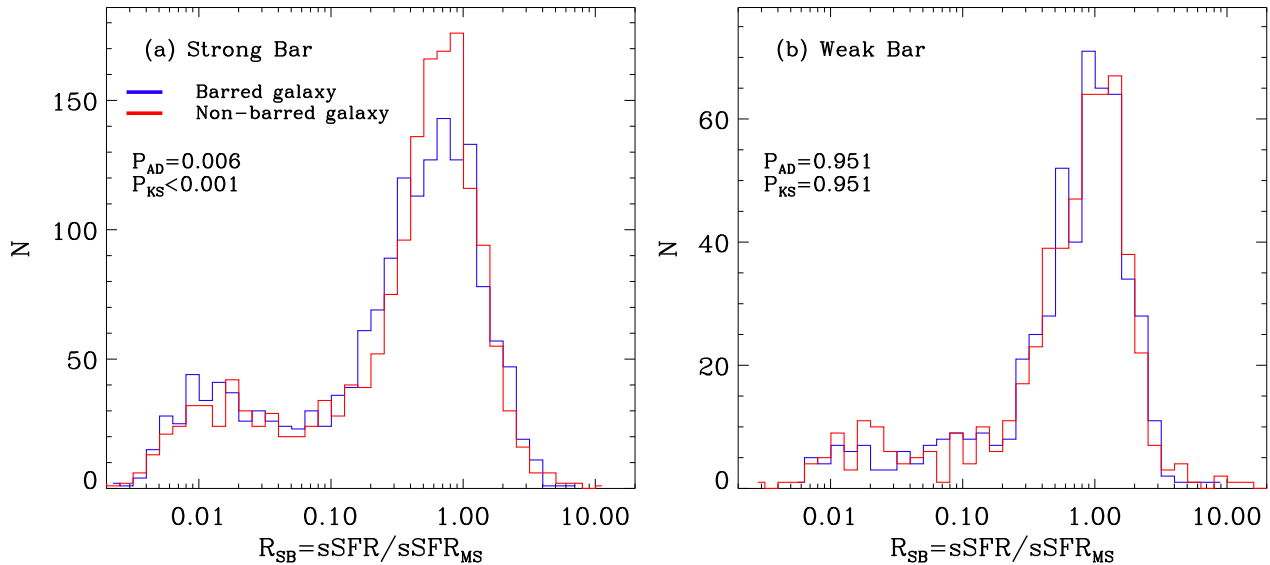}
\caption{Starburstiness R$_{SB}$ distributions of strongly (left) and weakly (right)
  barred galaxies and their control samples. 
Blue and red lines represent barred and non-barred galaxies, respectively.
}\label{fig-rsb}
\end{figure*}
\begin{figure}
\center
\includegraphics[width=0.48\textwidth]{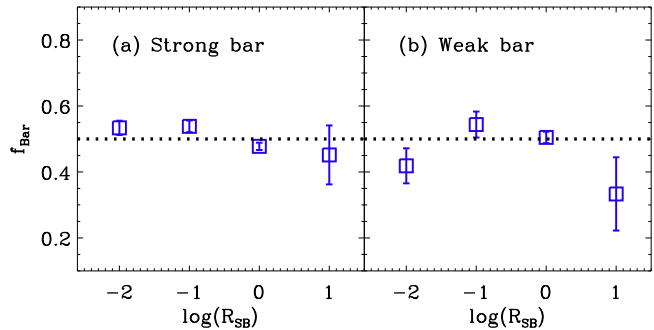}
\caption{Barred galaxy fraction as a function of starburstiness, log(R$_{SB}$):
  (left) strongly barred galaxies, (right) weakly barred galaxies.
The dashed line represents $f_{\rm Bar}=0.5$.
}\label{fig-rsbfrac}
\end{figure}

\begin{figure*}[t]
\center
\begin{tabular}{cc}
\includegraphics[width=0.49\textwidth]{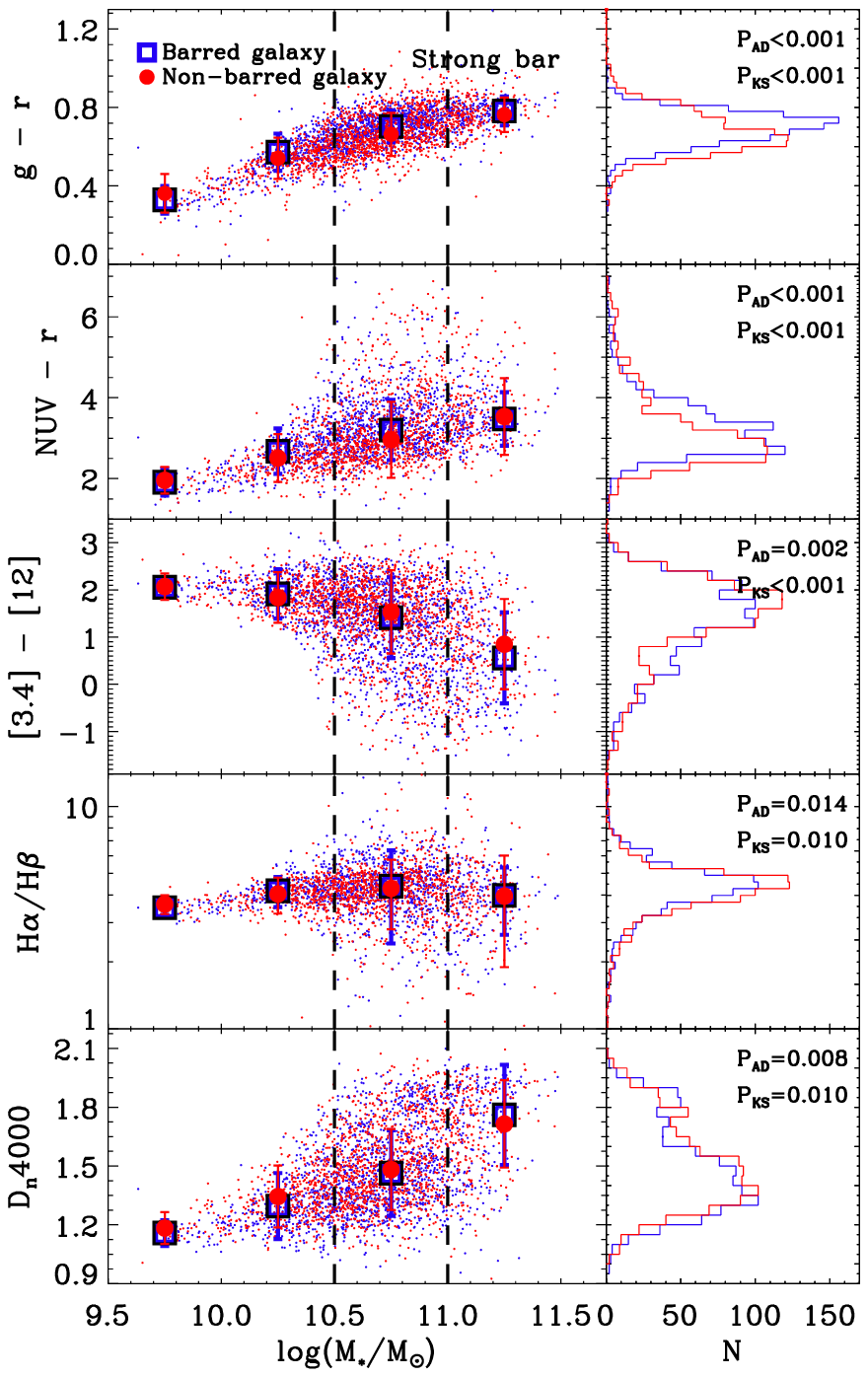}
\includegraphics[width=0.49\textwidth]{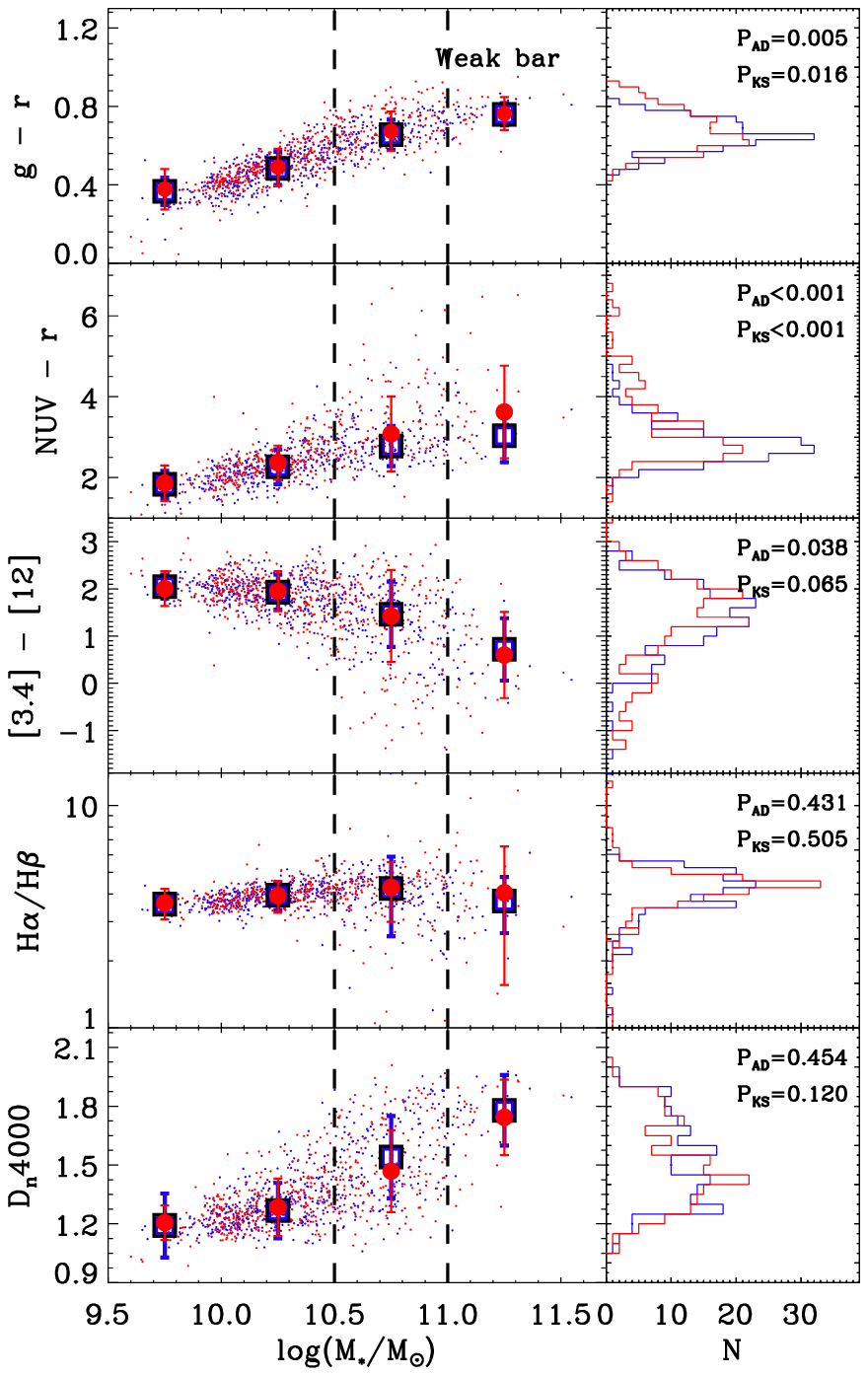}
\end{tabular}
\caption{(Left) Optical, near-UV, mid-IR colors, Balmer decrement (H$_\alpha$/H$_\beta$)
  and $D_{n}$4000 (from top to bottom)
  as a function of stellar mass for strongly barred galaxies (blue dots)
  and their control sample (red dots).
Large blue open square and red filled circle
  are median values for barred and non-barred galaxies at each mass bin, respectively. 
The histogram in each panel shows the distribution of each parameter
  for barred (blue line) and non-barred (red line) 
  galaxies in a narrow mass range 10.5 $\le$ log$(M_*/M_\odot)$ $<$ 11.0.
(Right) Same as left panels, but for weakly barred galaxies and their control sample.
}\label{fig-color}
\end{figure*}

\section{RESULTS}\label{results}

\subsection{Star Formation Activity of Barred Galaxies}\label{sfa}

The top panels of Figure \ref{fig-sfrm} show the SFRs of barred (blue dots) 
  and non-barred (red dots) galaxies as a function of stellar mass.
The left and right panels display strongly and weakly barred galaxies, respectively.
We also show the locus of main-sequence star-forming galaxies
  defined in \citet{elbaz07}.
The solid line is the best-fit to the SDSS main sequence star-forming galaxies,
  and the upper and lower dashed lines are a factor 4 above and below this fit.
We adopt the main-sequence locus of \citet{elbaz07} 
  for ease of comparison with other studies,
  and note that the following results do not change much even though we
  newly define the main sequence using the galaxies in this study.
  
We then consider the galaxies within these dashed lines as main sequence (MS),
  those above the upper dashed line as starburst (SB, ${\rm SFR>4\times SFR_{MS}}$), 
  and those below the lower dashed line as quiescent galaxies 
  (QS, i.e. ${\rm SFR<0.25\times SFR_{MS}}$).
The majority of our samples of barred and non-barred galaxies
   are in the main sequence as expected.
It is difficult to tell the difference in the distributions of barred
  and non-barred galaxies in this SFR-M$_*$ plane by eye,
  consistent with previous results \citep{willett15}.
Weakly barred galaxies tend to have more low-mass galaxies
  with $M_*<3\times10^{10}$ M$_\odot$ than high-mass galaxies,
  which is different from the case of strongly barred galaxies.  
The bottom panels show the specific SFRs (sSFRs) for the same samples 
  as a function of stellar mass. 
Both strongly and weakly barred galaxies follow the tight star-forming sequence.

We calculate the fraction of galaxies in each group of star-formation mode, 
  and summarize the result in Table \ref{tab:frac}.
The fraction of main-sequence galaxies is very high 
  for both strongly and weakly barred galaxies as expected,
  but is lower in strongly barred galaxies than in weakly barred galaxies.
The left columns for strongly barred galaxies show
  that the fractions of starburst and main sequence groups 
  are smaller in barred galaxies than in non-barred galaxies; 
  the fraction of quiescent galaxies for strongly barred galaxies 
  is larger than for non-barred galaxies accordingly.
The error represents 68\% ($1\sigma$) confidence interval 
  that is determined by the bootstrap resampling method.
Considering the errors, the difference between 
  strongly barred galaxies and their control sample is not negligible.
However, the fraction of each group for weakly barred galaxies
  is similar to that for their control sample.
    
To better compare the star formation activity between
  barred and non-barred galaxies by minimizing the mass effects, 
  we plot the starburstiness ($\rm R_{SB}$) distribution in Figure \ref{fig-rsb}. 
The starburstiness is a measure of the excess in sSFR of a galaxy
  compared to sSFR of a main-sequence star-forming galaxy with the same mass,
  and is defined by ${\rm R_{SB} = sSFR/sSFR_{MS}}$ \citep{elbaz11}. 
Figure \ref{fig-rsb} displays the starburstiness distributions
  of strongly (left panel) and weakly (right panel) barred galaxies. 
The left panel shows that the strongly barred galaxies (blue histogram) have a wider, 
  lower peak than their control sample (red histogram) at $\rm R_{SB}\approx$ 0.7. 
The K-S and A-D k-sample tests on the starburstiness distributions
  of strongly barred galaxies and their control sample
  yield $p$-values of $p_{\rm KS}<0.001$ and $p_{\rm AD}<0.006$,
  indicating a significant difference between the two distributions.
However, the right panel for weakly barred galaxies shows no such difference between
  barred (blue histogram) and non-barred (red histogram) galaxies,
  confirmed by the K-S and A-D k-sample tests.
  
To explore a possible mass dependence of the starburstiness distribution, 
  we also examine the starburstiness distributions at several narrow mass ranges 
  (not shown here, but from 
  log($M_*/M_\odot$)$=9.5$ to log($M_*/M_\odot$)$=11$ with a 0.5 dex bin). 
The histogram at each mass bin shows the similar results to the one using all the galaxies 
  (i.e. statistically different distribution between strongly-barred galaxies and 
  their control sample except the mass range of $10.0<$log($M_*/M_\odot$)$<10.5$, 
  and no statistically difference between weakly-barred galaxies and their control sample), 
  suggesting that the starburstness difference between barred and non-barred galaxies 
  persists throughout the entire mass range.

When we compare $\rm R_{SB}$ distributions 
  between strongly and weakly barred galaxies,
  the strongly barred galaxies appear to 
  have a relatively higher fraction of quiescent galaxies
  (R$_{\rm SB}$ $\approx$ 0.01) than 
  the weakly barred galaxies (see also Table \ref{tab:frac}).
The K-S and A-D k-sample tests on the starburstiness distributions
  of strongly and weakly barred galaxies reject the null hypothesis
  at $>$3$\sigma$ level,
  suggesting a significant difference in the star formation activity between the two.

To better understand the correlation between the presence of bars 
  and the starburstiness of galaxies,
  we show the fraction of barred galaxies
  as a function of starburstiness in Figure \ref{fig-rsbfrac}.
The left panel for the strongly barred galaxies 
  shows a hint of decrease of bar fraction with starburstiness, 
  but it is not conclusive because of large error bars.
The weakly barred galaxies in the right panel show no clear dependence of bar fraction
  on the starburstiness.

\subsection{Comparison of Physical Parameters between Barred and Non-barred Galaxies}\label{comparison}
\subsubsection{Multiwavelength Data from near-UV to mid-infrared}

Figure \ref{fig-color} shows several multiwavelength colors
  of barred and non-barred galaxies as a function of stellar mass
  (left: strongly barred galaxies, right: weakly barred galaxies).
Beginning from the top panel, we display, sequentially, 
  $\it{g-r}$, NUV$ - r$, mid-infrared [3.4]$-$[12] colors, 
  the flux ratios between H$\alpha$ and H$\beta$ (i.e., Balmer decrement), and
  $D_n4000$.
The optical $\it{g-r}$ color is a good tracer of star formation activity
  in galaxies (e.g., \citealt{strateva01,blanton03}).
The NUV$ - r$ and [3.4]$-$[12] colors are good indicators of recent
  star formation activity of galaxies with slightly different timescales
  (e.g., \citealt{ko13,ko16,lee15,lee17}); both NUV and mid-infrared are sensitive to 
  very recent ($<1$ Gyr) star formation, 
  but only the mid-infrared is sensitive to star formation 
  over longer (up to $\sim2$ Gyr) timescales.
The flux ratio between H$\alpha$ and H$\beta$ (i.e., Balmer decrement)
  is a measure of dust extinction.
When there is no dust in galaxies,
 H$\alpha/$H$\beta$ ratios are expected to be 2.86 and 3.1
 for star-forming and AGN-host galaxies, respectively
  (in the nominal case B recombination
  for $T=10,000$ K and $n_e\approx 10$ cm$^{-3}$, \citealt{ost06}).
Therefore, flux ratios larger than these values indicate dust extinction.
$D_{n}4000$ is a measure of the 4000~\AA~break \citep{bruzual83,balogh99},
  which results from an accumulation of absorption lines of ionized metals 
  in low mass stars at wavelength $<$4000 \AA.
The amplitude of the break is smaller in galaxies with young stellar populations
  because the opacity decreases in hot young stars.
It is larger for old metal-rich populations.
Therefore, $D_n4000$ is a useful measure of the age of the stellar population.

\begin{figure*}[t]
\center
\includegraphics[width=0.9\textwidth]{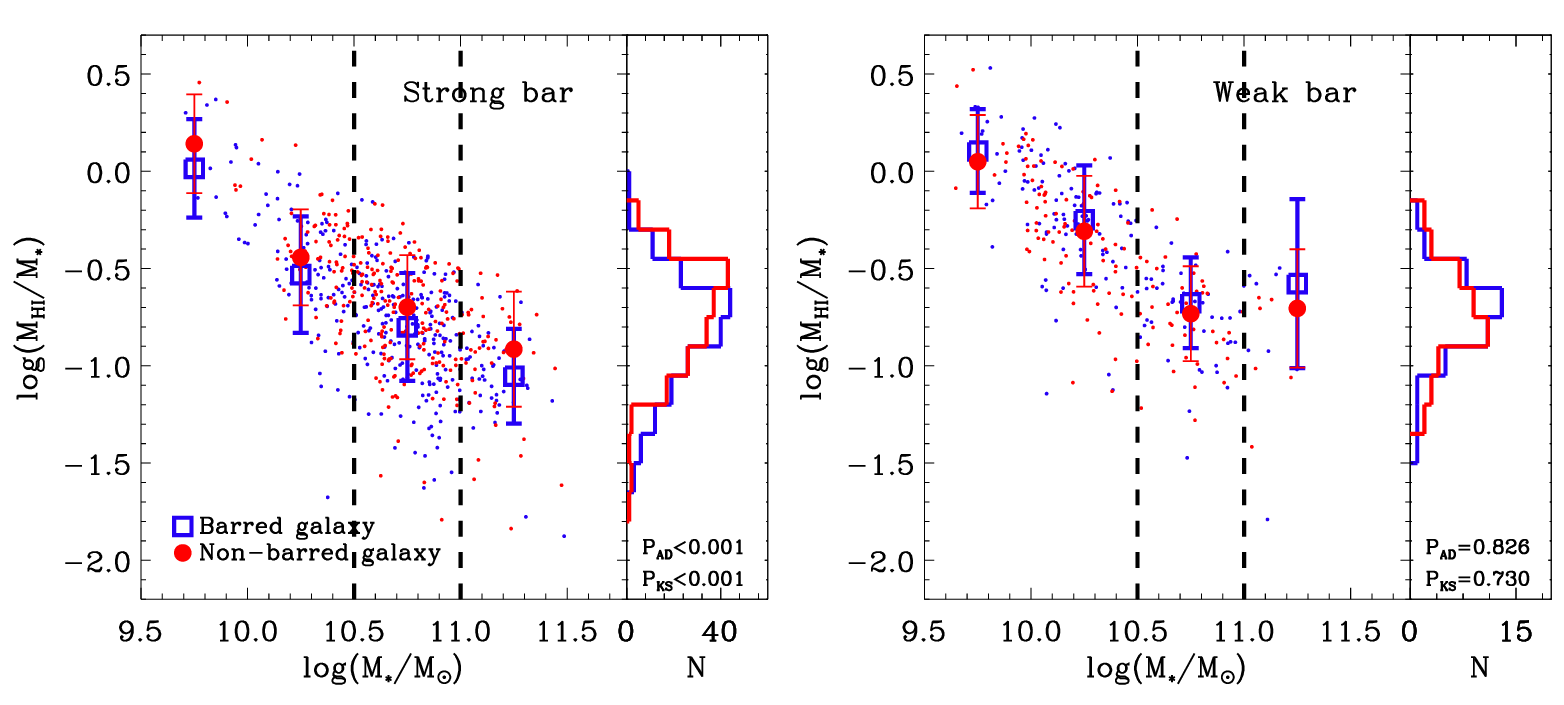}
\caption{(Left) Mass fraction of HI gas as a function of stellar mass
  for strongly barred galaxies and their control sample.
Blue and red dots show barred and non-barred galaxies, respectively. 
Blue open rectangles and red filled circles are median values at different mass bins.
The right panel shows the histograms of gas fraction for barred (blue line)
  and non-barred (red line) galaxies.
(Right) Same as left panels, but for weakly barred galaxies and their control sample.
}\label{fig-HI}
\end{figure*}

\begin{figure*}
\center
\includegraphics[width=0.9\textwidth]{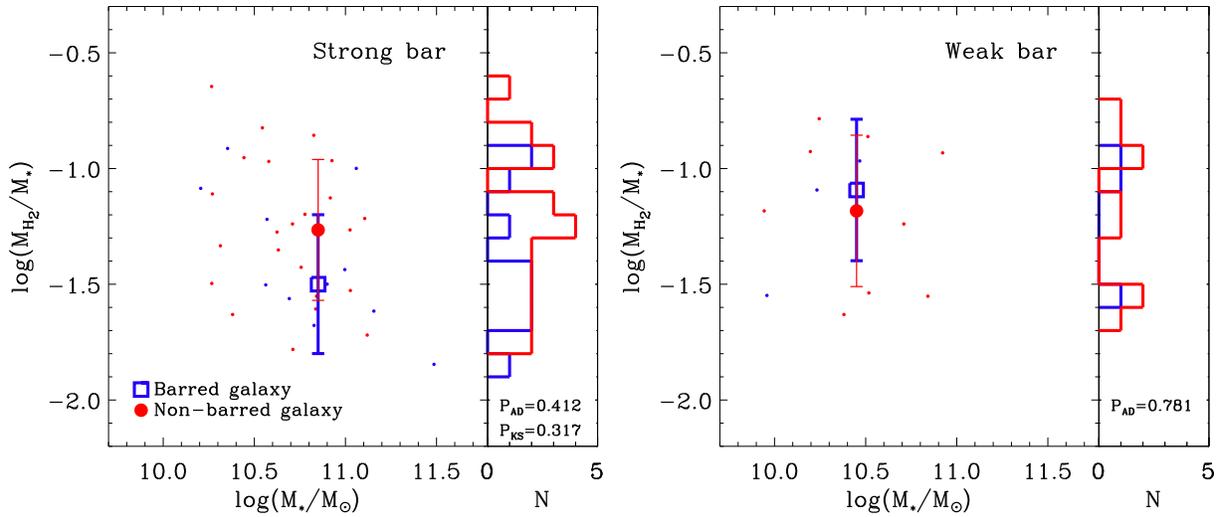}
\caption{Same as Figure \ref{fig-HI}, but for 
  H$_{2}$ gas mass fraction. 
}\label{fig-H2}
\end{figure*}

\begin{figure*}
\center
\includegraphics[width=0.9\textwidth]{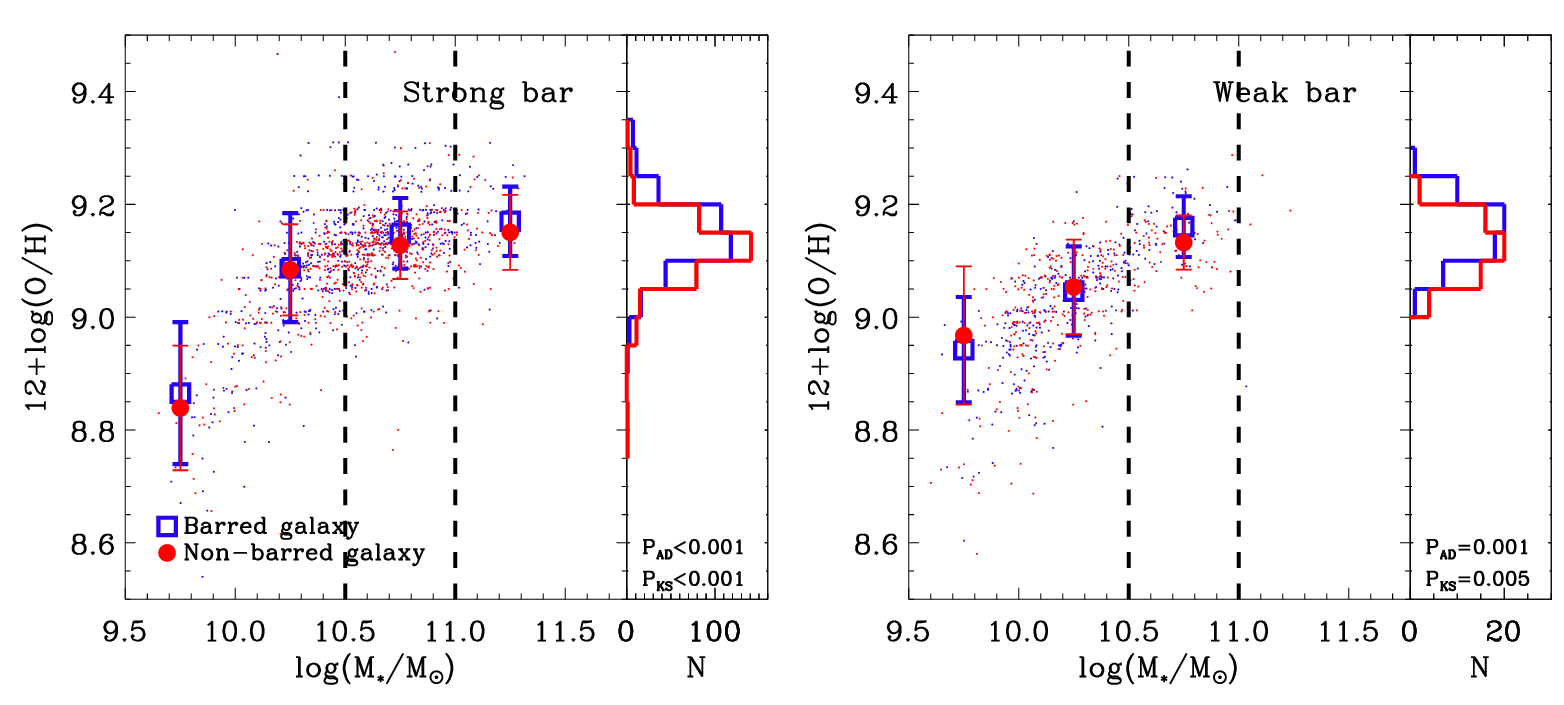}
\caption{Same as Figure \ref{fig-HI}, but for 
  gas metallicity.
}\label{fig-metal}
\end{figure*}

Figure \ref{fig-color} shows that as stellar mass increases,
  $\it{g-r}$ and NUV$-r$ colors increase \citep{blanton03,lee12b,ko13},
  [3.4]$-$[12] color decreases \citep{ko13},   and 
  $D_n4000$ increases \citep{geller14, geller16}.
The difference between barred and non-barred galaxies
  in each mass bin is not obvious.
To minimize the mass effects on the comparisons of color distributions
  between the two samples, we show the histogram of each parameter
  for the galaxies in a narrow mass bin 
  (i.e. 10.5 $\leq$ log(M$_{\star}$/M$_\odot$) $<$11.0),
  as indicated with vertical dashed lines in Figure \ref{fig-color}
  where we can have a large number of galaxies.
The strongly barred galaxies tend to be 
  redder in $\it{g-r}$ and NUV$-r$ colors, and
  bluer in [3.4]$-$[12] color than their control sample.
This systematic difference in colors between the strongly barred galaxies
  and their control sample seems to result from the
  different dust extinction between the two.
The second panel from the bottom
  shows that there is a statistically significant difference
  in the Balmer decrement distribution
  between strongly-barred galaxies and their control sample;
  barred galaxies appear to experience slightly more dust extinction than non-barred galaxies,
  which can make the barred galaxies redder in $\it{g-r}$ and NUV$-r$ colors.
There also could be an additional effect on this color difference, 
  which is hinted by the smaller number of actively star-forming galaxies 
  in the sample of strongly barred galaxies than in their control sample in Figure \ref{fig-rsb}.
  
On the other hand, $D_n4000$ does not show a systematic difference
  between strongly barred galaxies and their control sample.
Although the K-S and A-D k-sample tests on $D_n4000$ distribution
  at 10.5 $\leq$ log(M$_{\star}$/M$_\odot$) $<$11.0
  indicate a significant difference between the two,
  the Student’s t-test suggests that the probability that the two samples 
  have significantly different means is only at $<2\sigma$ level;
  this means that the low $p$-values of the K-S and A-D k-sample tests 
  are not because of different means (or medians)
  but because of different distributions.
The right panels for the weakly barred galaxies show that
  barred and non-barred galaxies are statistically different
  only in the NUV$-r$ color distribution.

\subsubsection{Gas Properties}

Star formation activity is directly connected to the amount of gas in galaxies.
We therefore compare the amounts of atomic (HI) and molecular (H$_{2}$) gas
  between barred and non-barred galaxies.
Figure \ref{fig-HI} shows the fraction of HI gas mass as a function of stellar mass
  for our samples.
As expected, the HI gas fraction decreases with stellar mass.
The left panel for strongly barred galaxies shows that
 the HI gas fraction of barred galaxies is systematically lower than 
 that of non-barred galaxies,
 consistent with the results in previous studies \citep{masters12,sodi17}.
The K-S and A-D k-sample tests on the HI gas fraction distributions
  between the two also reject the null hypothesis that
  the two distributions are extracted from the same parent population
  at $>$3$\sigma$ level.
Interestingly, the weakly barred galaxies in the right panel
  appear to have higher HI gas fractions 
  than non-barred galaxies in all mass bins, 
  different from the case of strongly barred galaxies.
However, the difference is not statistically significant.

We also show the fraction of H$_2$ gas mass as a function of stellar mass
  in Figure \ref{fig-H2}.
The left panel for strongly barred galaxies shows a hint of 
  different H$_2$ gas fractions between barred
  and non-barred galaxies (i.e. lower gas fraction in barred galaxies),
  similar to the result of HI gas fraction.
However, the difference is not statistically significant
  as the K-S and A-D tests suggest.
The right panel for the weakly barred galaxies
  does not show any meaningful comparison 
  between barred and non-barred galaxies
  because of small number statistics.

We also examine the distribution of gas depletion time,
  $t_{\rm dep} \equiv M_{\rm gas}/{\rm SFR}$ for both HI and H$_2$ gas,
  but do not find any significant difference between
  barred and non-barred galaxies 
  for both strongly and weakly barred galaxies (not shown here).

Figure \ref{fig-metal} shows a comparison of gas metallicity
  between barred and non-barred galaxies.
The metallicity of galaxies increases with stellar mass, 
  which follows a well-known mass-metallicity relation
  \citep{lequeux79,tremonti04,zahid13}.
The metallicity of strongly barred galaxies is systematically higher
  than of their control sample at each mass bin.
This is confirmed by the K-S and A-D k-sample tests
  at $>3\sigma$ level, consistent with the results
  of \citet{vera16}.
Overall gas metallicity between weakly-barred galaxies and their control sample
  is similar (confirmed by the K-S and A-D k-sample tests), 
  but the galaxies at 10.5 $\leq$ log(M$_{\star}$/M$_\odot$) $<$11.0
  show a $>2.8\sigma$ difference.

\subsection{Stellar Populations of Barred Galaxies: Fit to the Optical Spectra with STARLIGHT}\label{sfh}

\begin{figure*}
\center
\includegraphics[width=0.8\textwidth]{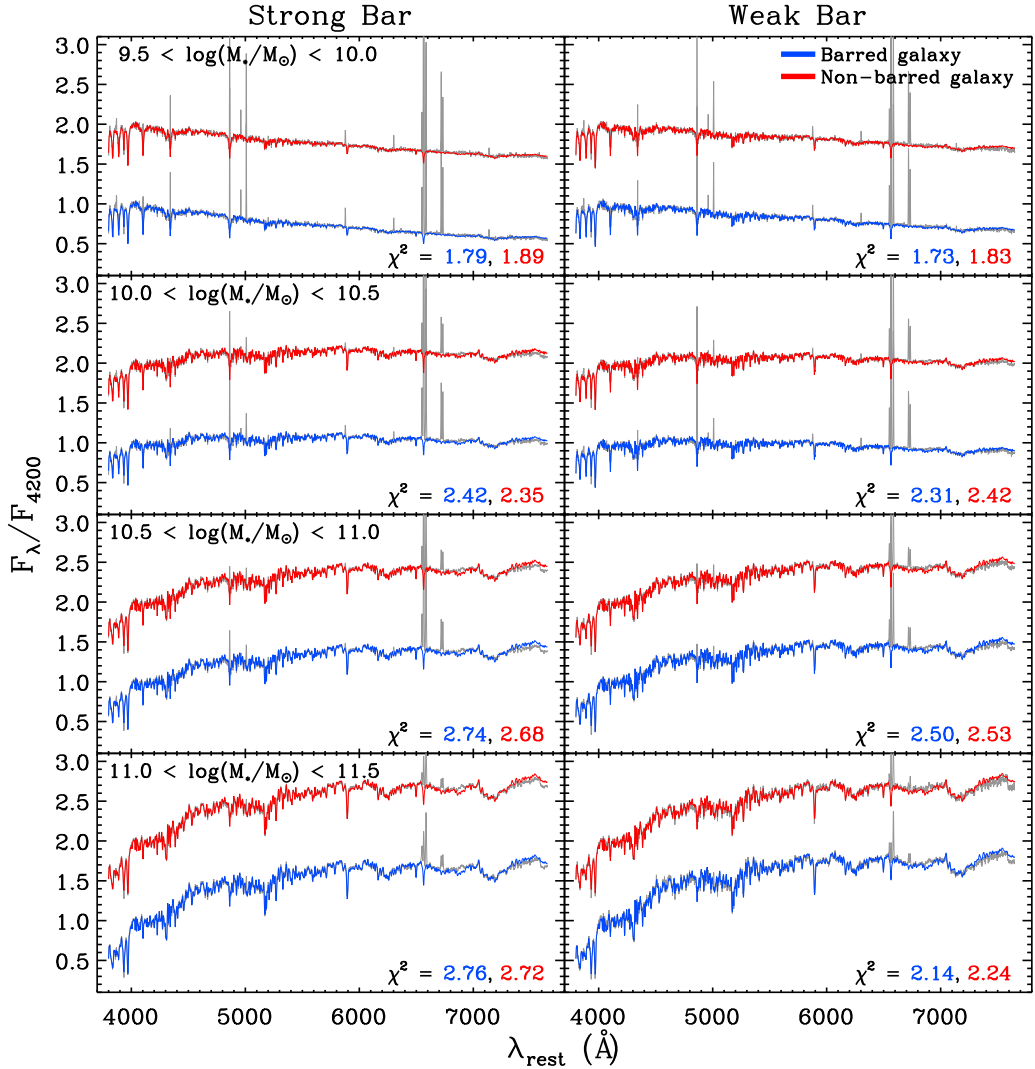}
\caption{(Left) STARLIGHT fit to the stacked spectra of 
  strongly barred galaxies and their control sample at different stellar mass bins.
We shift the spectra of non-barred galaxies by adding unity for clarity.
The gray and color-coded lines represent the stacked spectra and 
  the best-fit models, respectively. 
Lines are not clearly visible because of overlap.
(Right) Same as for left panels, but for weakly barred galaxies and their control sample.
}\label{fig-spec}
\end{figure*}

In this section, we compare stellar populations of barred and non-barred galaxies.
We perform a decomposition of stellar populations of barred and non-barred galaxies
  by fitting the SDSS optical spectra with the spectral fitting code, STARLIGHT \citep{cid05}.
To minimize the mass effects on the stellar population comparison and
  to increase the signal-to-noise ratio of the spectra, 
  we divide the galaxy samples into four mass bins and 
  stack the rest-frame spectra in each mass bin. 
We normalize the individual spectrum at rest-frame 4150$-$4250~\AA~ and
  take the median for the stacking.
The wavelength coverage for the stacked spectra is $3800-7650$~\AA.
We fit to the spectra using 45 spectral templates of \cite{bc03} with 
  15 ages (between 1 Myr and 13 Gyr) and 3 metallicities ($Z=$0.004, 0.02, 0.05).
These templates are generated from STELIB library \citep{leborgne03}
  with Padova evolutionary tracks \cite{bertelli94} and \citet{chabrier03} IMF.
We perform the STARLIGHT fit 100 times with different seeds 
  for the random number generator 
  and adopt the medians of derived parameters as the best-fit results. 
Figure \ref{fig-spec} shows the stacked spectra of galaxies in different mass bins
  (gray lines) with the best-fit STARLIGHT models for barred (blue lines) and 
  non-barred (red lines) galaxies.

\begin{figure*}
\center
\includegraphics[width=0.9\textwidth]{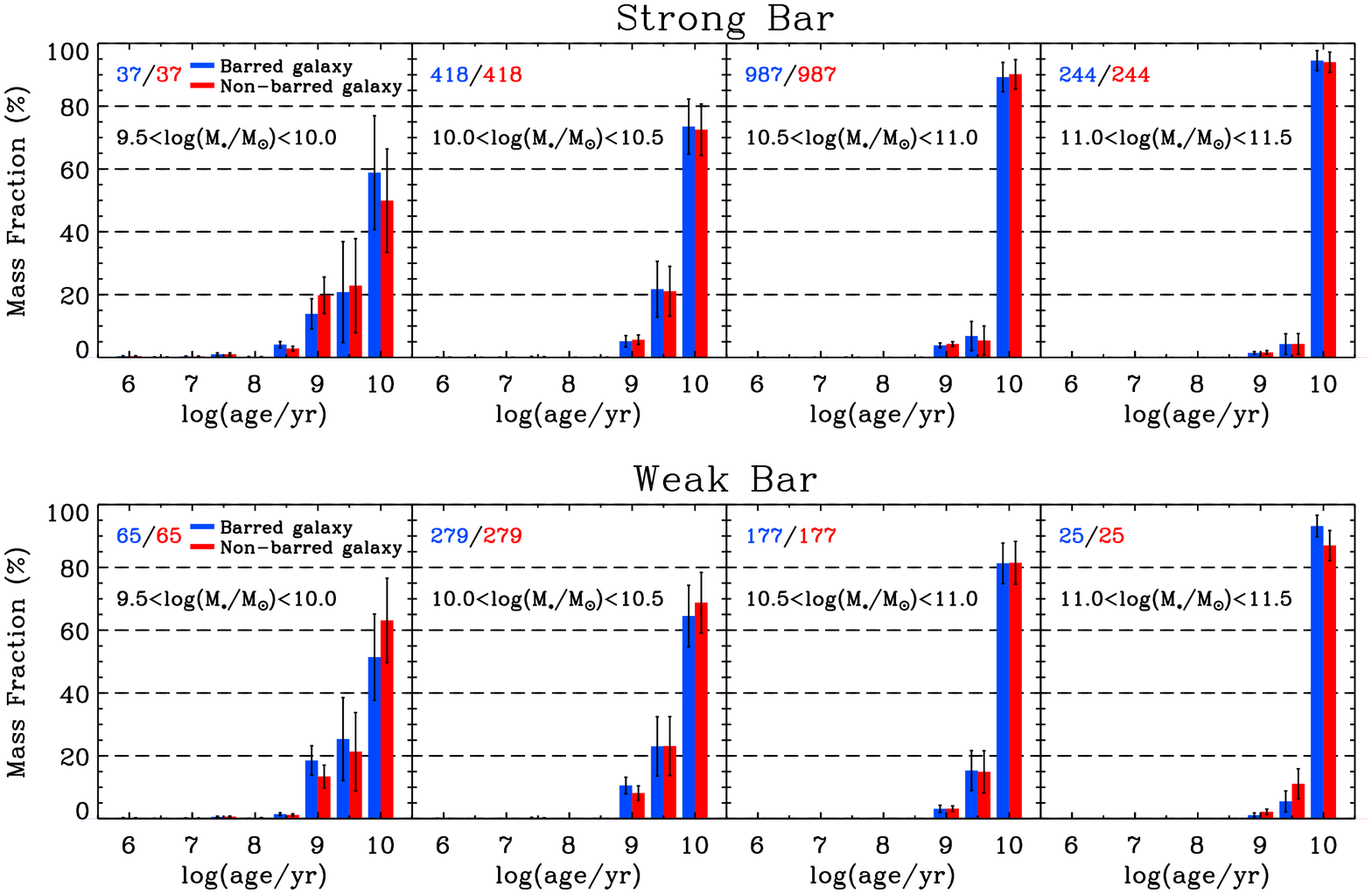}
\caption{(Top) The fractions of stellar populations with different ages 
  in the total stellar mass for strongly barred galaxies (blue histograms) and 
  their control sample (red histograms) at different mass bins. 
Error bars indicate 1$\sigma$ distributions from the 100 fits with different seeds.
(Bottom) Same as top panels, but for weakly barred galaxies and their control sample.
}\label{fig-mfrac}
\end{figure*}

The top panels of Figure \ref{fig-mfrac} show the resulting mass fractions 
  of different stellar populations for strongly barred galaxies and their control sample.
As expected, the mass fractions of young ($\leq$2 Gyr) and 
  intermediate-age ($\sim$2--5 Gyr) populations decrease with stellar mass 
  for both barred and non-barred galaxies. 
The difference in the mass fraction 
  between barred and non-barred galaxies is not obvious.
The weakly barred galaxies in the bottom panels show similar trends;
  the mass fraction of old stellar populations increases with stellar mass, 
  and the difference between barred and non-barred galaxies is not significant.

\begin{figure*}
\center
\includegraphics[width=0.8\textwidth]{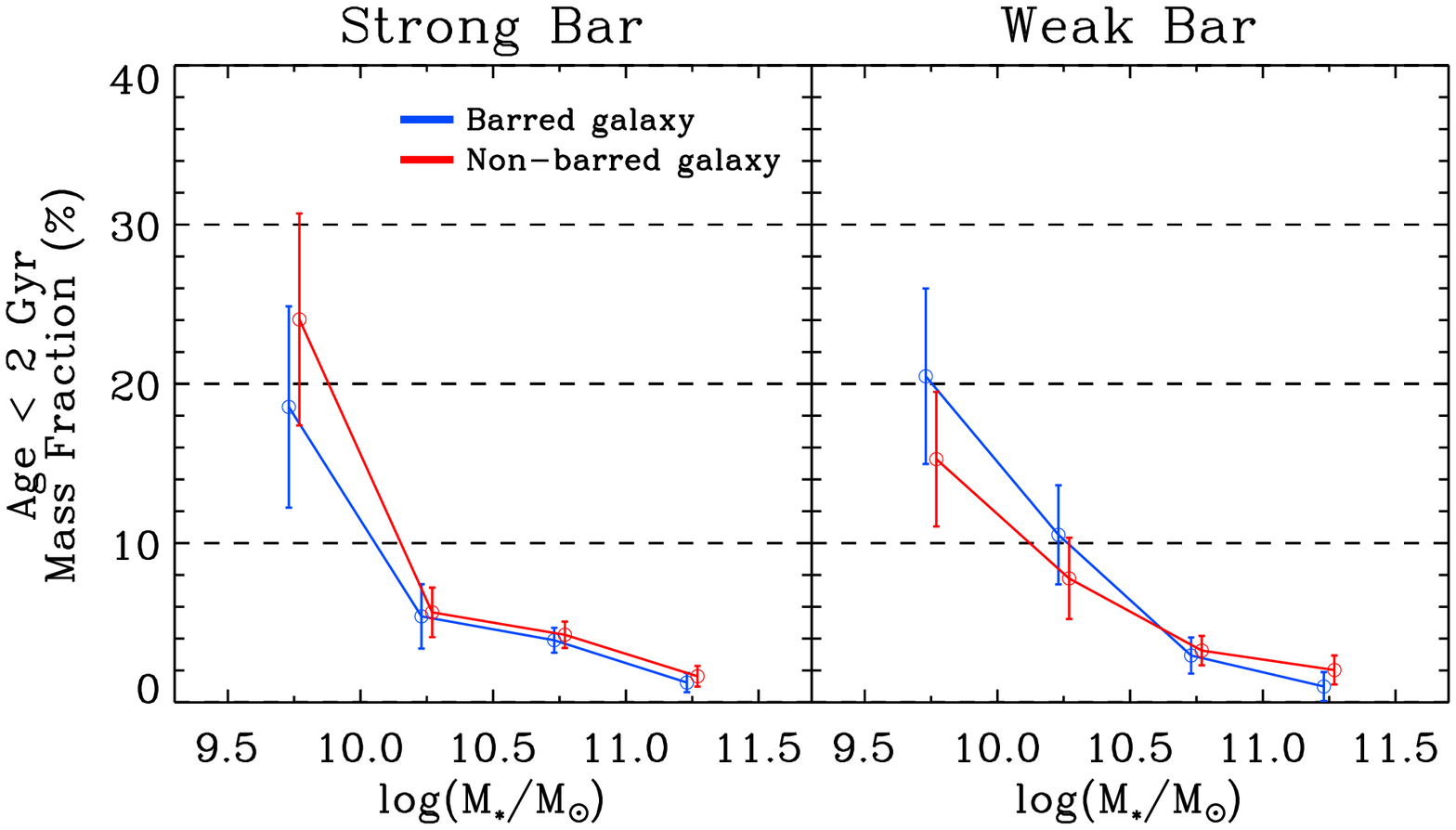}
\caption{(Left) The mass fractions of young stellar populations ($<$ 2 Gyr) 
  as a function of stellar mass for strongly barred galaxies (blue circles) and 
  their control sample (red circles).
(Right) Same as left panels, but for weakly barred galaxies and their control sample.
}\label{fig-ypop}
\end{figure*}

To make a more quantitative comparison between barred and non-barred galaxies,
  we plot the mass fraction of young stellar population with age $<$ 2 Gyrs
  in Figure \ref{fig-ypop}.
Both barred and non-barred galaxies show that the mass fraction of young stellar population
  decreases with stellar mass, 
  which confirms the visual impression of Figure \ref{fig-mfrac}.
At the lowest mass bin (9.5$<$log(M$_*$/M$_\odot$)$<$10.0),
  the strongly barred galaxies show a possible hint of lower mass fraction of 
  young stellar population than in non-barred galaxies,
  but it is not statistically significant.
The weakly barred galaxies in the right panel 
  show a pattern different from strongly barred galaxies
  (i.e. higher mass fraction of young stellar population in barred galaxies
  than in non-barred galaxies), but again it is not statistically significant.
\citet{ko16} examined the dependence of the fit 
  on the choice of spectral templates with various combinations of 
  age/metallicity distributions (i.e. different star formation history), 
  stellar population models, and IMFs.
They found that the mass fraction of young and intermediate-age stars does not change much 
  with different combinations of age and metallicity distributions,
  which suggests that our results would not change much 
  with different choices of spectral templates for the fit.


\section{DISCUSSION}\label{discuss}

We use various tracers of star formation activity in galaxies
  to examine the difference between barred and non-barred galaxies.
The comparisons of starburstiness,
   $\it{g-r}$, NUV$-r$, mid-infrared [3.4]$-$[12] colors,  
   the HI gas fraction, and gas metallicity
   between strongly barred galaxies and their control sample show
   significant differences between the two;
   the barred galaxies generally show weaker star formation activity
   than non-barred galaxies.
The H$_2$ gas fraction and the mass fraction of young stellar population ($<$ 2 Gyrs)
  from the decomposition of the optical spectra also show similar differences,
  but the statistical significance is not very high.
On the other hand, most star formation activity tracers
  for the weakly barred galaxies show 
  no such differences between barred and non-barred galaxies.

The weaker star formation activity of strongly barred galaxies than their control sample
  is consistent with previous results \citep{vera16}.
Because the weakly barred galaxies show no such difference, 
 comparisons between barred and non-barred galaxies 
  including both strongly and weakly barred galaxies 
  would have made the possible difference weak,
  which can result in no increase in star formation activity in barred galaxies
  \citep{pompea90, martinet97, chapelon99,cheung13,willett15}.

We also show that the HI gas fraction of strongly barred galaxies is, on average, lower than
  their control sample, confirming the previous findings \citep{sodi17} even though
  some studies did not distinguish strongly and weakly barred galaxies \citep{masters12}.
Figure \ref{fig-H2} shows a hint of lower H$_2$ gas fraction in barred galaxies than 
  in non-barred galaxies, 
  but it needs to be examined with more H$_2$ gas mass measurements of galaxies
  for better statistical significance.
We could not find a significant difference in the distribution of gas depletion time
  between barred and non-barred galaxies for both HI and H$_2$ gas.
Similarly, \citet{saintonge12} found only a marginal difference 
  in molecular gas depletion time
  between barred and non-barred galaxies.
\citet{sodi17} also found that only strongly barred galaxies
  show a mild increase of bar fraction with atomic gas depletion time.

Many numerical simulations showed that the star formation activity
  in central regions of galaxies could be enhanced by the presence of galactic bars
   \citep{shlosman90,athan94,combes01,kimss11}.
Although at first instance our results seem at odds with these theoretical expectations, 
  we explore two scenarios that can help to ease the discrepancy. 
First, \cite{carles16} suggested that
  the star formation activity of barred galaxies 
  could be triggered when the bars are formed and
  the SFRs of barred galaxies become similar to or lower than non-barred galaxies in 2 Gyrs;
  the difference between barred and non-barred galaxies
  would be larger strongly barred galaxies.
The consumption of hydrogen gas is accelerated along with 
  the star formation activity in barred galaxies, then the amount of gas
  in barred galaxies also becomes similar to or lower than for non-barred galaxies.
These could explain why many observations show no increase
  in {\it current} star formation activity of barred galaxies compared to
  non-barred galaxies.
These also could explain lower gas mass and higher metallicity currently in barred galaxies
  than in non-barred galaxies.  
This was the motivation of the decomposition of 
  the stellar populations of barred and non-barred galaxies
  by fitting to the SDSS spectra with STARLIGHT in this study.
However, our decomposition could not find any significant difference
  in the mass fraction of young stellar populations ($<2$ Gyrs)
  between barred and non-barred galaxies even though
  there is a hint of difference only at the lowest mass bin
  (9.5$<$log(M$_*$/M$_\odot$)$<$10.0).

Second, the star formation indicators we consider in this study are global galaxy parameters
  rather than central ones, and the bar-induced star formation activity is expected 
  to take place only in central regions where bars can shock the gas. 
Observational evidence of this localized star formation in the central region of barred galaxies 
  has already been reported.
For instance, by studying the central-to-total star formation activity in a large volume-limited sample 
  of SDSS galaxies, \citet{wang12} found that more than half of the galaxies 
  with enhanced central star formation activity have bars.
More recently, \citet{lin17} analyzed the integral field spectroscopic data of 57 galaxies, 
  and found that among the 17 ``turnover'' galaxies with rejuvenated pseudobulges 
  most of them (15/17) have bars.
More two-dimensional spectroscopic observations of these barred galaxies 
  will be useful for better localizing the star formation activity in barred galaxies.

If bar-triggered star formation is not reflected globally, 
  it would be difficult to imagine these systems being efficient 
  in exhausting their gas though enhanced star formation. 
In this case, the low gas fraction in barred galaxies could be explained 
  through the inhibiting effect that the gas has in the formation and growth of bars
  (i.e. bars form later and grow more slowly in gas-rich galaxies than in gas-poor galaxies),
  as shown in hydrodynamical simulations \citep{villa-vargas10,athan13}.
We could not distinguish which scenario is more plausible at this stage, 
  but studying barred galaxies at high redshift to understand their evolution 
  will be helpful for testing these scenarios.
 
Bar strength has been considered as one of important parameters 
  in the star formation activity of barred galaxies 
  \citep{athan03,buta05,nair10,hoyle11}.
In this study, we construct the control samples of non-barred galaxies
  for strongly and weakly barred galaxies separately
  by matching the stellar mass and redshift distributions. 
Our results suggest that many physical properties of strongly and weakly
  barred galaxies could be different;
  the current star formation activity of strongly barred galaxies appears
  lower than that of their control sample, but the weakly barred galaxies
  do not differ much from their control sample. 
If we assume that weak bars in galaxies evolve into strong bars
  and that the star formation activity is enhanced because of the presence of bars,
  we expect enhanced star formation activity in weakly barred galaxies
  compared to non-barred galaxies.
However, we found no significant difference 
  in star formation activity between weakly barred galaxies and 
  their control sample.
This can suggest that
  weakly and strongly barred galaxies are not linked by an evolutionary sequence or
  that the star formation activity of galaxies is not strongly affected 
  by the presence of bars.

We note that the barred galaxy sample used in this study 
  is constructed from the visual inspection of SDSS images 
  with a typical seeing $\sim$1.4 arcsec \citep{lee12a}. 
To identify bars in galaxies, the size of bars should be typically 3--4 times 
  larger than the point spread function (PSF) size. 
This means that some bars smaller than 2--5 kpc at 0.02 $\leq z \leq$ 0.05 
  (i.e. redshift range in our sample) could be missed in our galaxy sample because of seeing effect.
This can partly explain why some studies using nearby galaxy samples found 
  higher bar fractions than other studies that use relatively more distant 
  galaxy samples (e.g. \citealt{menendez07,diaz16}).
On the other hand, the size of bars increases with galaxy mass.
Then Figure 20 in \cite{diaz16} suggests that a significant amount of barred galaxies 
  less massive than a few times $10^{10}$ M$_\odot$ could be missed 
  in our galaxy sample because of the spatial resolution limit (i.e. $r_{\rm bar}\lesssim2$ kpc). 
This means that some less massive galaxies classified as non-barred galaxies in our sample 
  could indeed be barred galaxies. 
This can explain why some of the differences in physical parameters 
  between barred and non-barred galaxies are only prominent 
  for relatively more massive galaxies (e.g. 10.5$<$log(M$_*$/M$_\odot$)).

\section{CONCLUSIONS}\label{sum}

We use samples of strongly and weakly barred galaxies in the local universe,
  and compare their physical properties
  with those of non-barred galaxies focusing on the star formation activity.
Our primary results are:

\begin{enumerate}

\item The distributions of starburstiness (R$_{\rm SB}$), 
  a measure of the excess in sSFR of a galaxy
  compared to the sSFR of a main sequence star-forming galaxy,
  for strongly barred galaxies and their control sample are different;
  the strongly barred galaxies have a wider, lower peak near main sequence and
  have more quiescent galaxies than their control sample.
However, the starburstiness distribution of weakly barred galaxies
  is similar to that of their control sample.

\item The $\it{g-r}$, NUV$-r$, and mid-infrared [3.4]$-$[12] colors
  of strongly barred galaxies are statistically different from
  those of their control sample.
These color differences seem to result from the
  different dust extinction between the two,
  evidenced by the Balmer decrement (H$_\alpha$/H$_\beta$).
There also could be an additional effect on this color difference, which is that
  the star formation activity of strongly barred galaxies is,
  on average, lower than that of non-barred galaxies.
On the other hand, weakly barred galaxies do not show such significant differences
  in these multiwavelength parameters.
  
\item The HI gas fraction of strongly barred galaxies, on average, is lower
  than the one of their control sample.
There is also a hint of different H$_2$ gas fraction between 
  strongly barred galaxies and their control sample,
  which needs to be confirmed with more data.
Gas metallicity of strongly barred galaxies is, on average, higher
  than for their control sample.
Again, weakly barred galaxies do not show such significant differences
  in these gas properties.

\item The stellar population analysis of the optical spectra
 shows no significant difference between barred and non-barred galaxies.
However, strongly barred galaxies show a possible hint of lower mass fraction of 
  a young stellar population than in non-barred galaxies
  only at the lowest mass bin (9.5$<$log(M$_*$/M$_\odot$)$<$10.0).
Interestingly, weakly barred galaxies show a hint of higher mass fraction of 
  a young stellar population than their control sample in the same mass bin.

\end{enumerate}

Our results appear consistent with the idea that
  the star formation activity of strongly barred galaxies was
  enhanced in the past and is currently low (e.g. \citealt{carles16}).
Because of large gas consumption along with the star formation in the past, 
  the amount of gas and the gas metallicity of strongly barred galaxies
  are also expected to be currently low and high, respectively,
  consistent with our results.
This star formation history depends on the strength of bars.
To better understand the star formation activity in barred galaxies,
  a systematic survey of barred galaxies with
  two-dimensional spectroscopy 
  and studying high-redshift barred galaxies will be helpful.

\acknowledgments
We thank the anonymous referee for helpful comments. 
SSK and EK were supported by the National Research Foundation grant funded by the Ministry of Science, ICT and Future Planning of Korea (NRF-2014R1A2A1A11052367).
BCS acknowledges financial support through PAPIIT project IA103517 from DGAPA-UNAM.
Funding for the SDSS and SDSS-II has been provided by the Alfred P. Sloan Foundation, the Participating Institutions, the National Science Foundation, the U.S. Department of Energy, the National Aeronautics and Space Administration, the Japanese Monbukagakusho, the Max Planck Society, and the Higher Education Funding Council for England. The SDSS Web Site is http://www.sdss.org/.
The SDSS is managed by the Astrophysical Research Consortium for the Participating Institutions. The Participating Institutions are the American Museum of Natural History, Astrophysical Institute Potsdam, University of Basel, University of Cambridge, Case Western Reserve University, University of Chicago, Drexel University, Fermilab, the Institute for Advanced Study, the Japan Participation Group, Johns Hopkins University, the Joint Institute for Nuclear Astrophysics, the Kavli Institute for Particle Astrophysics and Cosmology, the Korean Scientist Group, the Chinese Academy of Sciences (LAMOST), Los Alamos National Laboratory, the Max-Planck-Institute for Astronomy (MPIA), the Max-Planck-Institute for Astrophysics (MPA), New Mexico State University, Ohio State University, University of Pittsburgh, University of Portsmouth, Princeton University, the United States Naval Observatory, and the University of Washington.
This publication makes use of data products from the {\it Wide-field Infrared Survey Explorer}, 
which is a joint project of the University of California, Los Angeles, 
and the Jet Propulsion Laboratory/California Institute of Technology, 
funded by the National Aeronautics and Space Administration.

\bibliographystyle{apj} 

\bibliographystyle{apj} 

{}

\end{document}